\documentclass[aps, pra, preprint, groupedaddress, amsfonts,
               amsmath, amssymb, showpacs, nofootinbib]{revtex4-1}

\usepackage{microtype}
\usepackage{graphicx}
\usepackage{amssymb}
\usepackage{mathtools}
\usepackage{epstopdf}
\usepackage[utf8]{inputenc}
\usepackage[T1]{fontenc}
\usepackage[usenames,dvipsnames]{xcolor}
\usepackage{hyperref}
\usepackage{subcaption}
\usepackage{mathptmx}
\usepackage{booktabs}
\usepackage[utf8]{inputenc}
\usepackage[T1]{fontenc}

\newcommand{\bra}[1]{\langle #1|}
\newcommand{\ket}[1]{|#1\rangle}
\newcommand{\braket}[2]{\langle #1|#2\rangle}

\usepackage{color}

\begin{document}
\title{Electron Emission in Antiproton-Hydrogen
Interactions Studied with the One-Centre Basis Generator
Method}

\author{Jay Jay Tsui}  
\email[]{jayjay93@yorku.ca}
\affiliation{Department of Physics and Astronomy, York University, Toronto, Ontario, Canada M3J 1P3}

\author{Tom Kirchner}  
\email[]{tomk@yorku.ca}
\affiliation{Department of Physics and Astronomy, York University, Toronto, Ontario, Canada M3J 1P3}
\date{\today}
\begin{abstract}
Electron emission from hydrogen atoms induced by antiproton impact at intermediate energies is investigated using the one-centre Basis Generator Method within a semi-classical impact-parameter framework. The formulation employs a single-centre expansion of the time-dependent Schrödinger equation with a pseudostate basis consisting of hydrogenic orbitals acted upon by powers of a Yukawa-regularized potential, providing a compact and effective representation of the electronic continuum.

Ionization probabilities are obtained by projecting the time-evolved wavefunction onto Coulomb continuum states, from which energy-differential cross sections (EDCS) are extracted. Exponential piecewise functions are constructed to interpolate between the pseudostate eigenenergies, yielding smooth EDCS profiles for each partial wave. The total EDCS, reconstructed by summing over all partial-wave contributions, exhibits good agreement with results from other pseudostate-based approaches.

\end{abstract}

\maketitle
\section{Introduction}
\label{sec:Intro}

Ion–atom collisions are fundamental processes in atomic and molecular physics, enabling detailed investigations into quantum-mechanical phenomena such as ionization, excitation, elastic scattering, and electron transfer. These interactions serve as critical testbeds for validating theoretical models but remain computationally demanding due to the complexity of the few-body Coulomb problem.

Among such systems, antiproton–hydrogen collisions have attracted extensive theoretical interest (see e.g., Refs. \cite{Schultz1989, Schultz1996, Sidky1998, Pons1999, Igarashi2000, Pons2000, Sakimoto2000, Azuma2001, Tong2001, Toshima2001, Jones2002, Voitkiv2003, Sahoo2004, McGovern2009, McGovern2010, Abdurakhmanov2011a, Abdurakhmanov2011b, Kirchner2011, Abdurakhmanov2016} and the review paper \cite{Kirchner2011} that summarizes the state of affairs up to 2011). The antiproton’s negative charge and large mass eliminate the possibility of electron capture and renders projectile deflection negligible. This allows it to be treated as a classical particle moving along a straight-line trajectory, simplifying the interaction dynamics to electron excitation and ionization. At intermediate impact energies, both protonium formation and relativistic effects can be neglected \cite{Schultz1996}, further reducing the complexity.

Under these conditions, the system reduces to a one-electron time-dependent Schrödinger equation (TDSE) for an electron moving in the combined Coulomb field of a stationary proton and a moving antiproton. The relevant ionization reaction is
\[
\bar{p} + \mathrm{H} \rightarrow \bar{p} + p + e^{-}
\]

The development of theoretical models for antiproton–hydrogen ionization includes pseudostate-based projection methods, which allow extraction of physical observables from coupled-channel calculations. As one example of such an approach, McGovern \textit{et al.}\ \cite{McGovern2009} introduced the coupled-pseudostate (CP) method within a semi-classical impact-parameter framework. In this formulation, the electronic wavefunction is expanded in a Laguerre-type pseudostate basis and evolved along classical projectile trajectories. Differential electron emission is obtained by projecting the time-evolved pseudostate wavefunction onto continuum Coulomb states, and fully differential ionization cross sections are extracted using the well-known connection between the impact-parameter and the wave treatments of the projectile motion.

Building upon this framework, McGovern \textit{et al.}\ later proposed a relaxed version of their method \cite{McGovern2010}, modifying the coupling scheme and improving the numerical projection procedure. This refinement enhanced the method’s stability and computational efficiency and broadened its applicability across a wider range of energies.

In parallel, Abdurakhmanov \textit{et al.}\ developed a Convergent Close-Coupling (CCC) fully quantal integral-equation approach formulated in the impact-parameter representation \cite{Abdurakhmanov2011a}, treating the three-body problem without decoupling the antiproton trajectory from the electronic dynamics. This method preserves full quantum coherence across the antiproton–hydrogen system.

Subsequently, they applied the CCC method to the same system \cite{Abdurakhmanov2011b}, introducing two methods for the calculation of energy-differential cross sections (EDCS): a so-called integration method which is based on projections onto Coulomb states and a simple summation method whose  ingredients are pseudostate populations and weight factors that involve differences of pseudostate energy eigenvalues.

More recently, a wave-packet extension of the CCC method (WP-CCC) has been introduced \cite{Abdurakhmanov2016}. This approach discretizes the continuum using localized wave packets rather than Laguerre-based pseudostates, mitigating basis linear-dependence issues and enabling high-resolution differential spectra.

A complementary route to modeling electronic dynamics in Coulombic systems is
provided by the Basis Generator Method (BGM)  \cite{Ludde1996, Kroneisen1999, Zapukhlyak2005}, also a pseudostate-based
approach but built on a different philosophy: Rather than aiming at completeness,
BGM basis sets are constructed with a view to merely spanning that subspace of Hilbert
space which is relevant for the system dynamics.
The BGM has been successfully applied to total cross section calculations for
a variety of collision systems, including bare-ion collisions with noble gas targets \cite{Kirchner1997, Kirchner1998, Keim2003, Keim2005, Henkel2009, Leung2015}, ion-molecule collisions \cite{Elizaga1999, Ludde2018}, and antiproton-impact collision systems \cite{Kirchner1998, Keim2005, Henkel2009, Ludde2021}. For the latter, energy loss calculations using Ehrenfest's theorem within a time-dependent density functional theory formulation were also carried out. More recently, the two-center BGM formulation has been extended to multicharged-ion collisions with atomic hydrogen \cite{Leung2019, Leung2022}. 

In this study, we apply the BGM to the calculation of \textit{differential} cross sections. We extract EDCS values at the pseudostate eigenenergies and implement a post-processing interpolation strategy to construct smooth, physically reliable cross-section curves. This approach offers a computationally efficient alternative to continuum integration or large-scale close-coupling schemes.

The remainder of this paper is organized as follows. Section~\ref{sec:theory} outlines the general theoretical framework, beginning with the TDSE in the semi-classical impact-parameter representation and introducing the so-called zero-overlap condition which represents a sufficient criterion for the stable extraction of EDCS at discrete energy points. Section~\ref{sec:ocbgm} describes the one-centre Basis Generator Method (OC-BGM) formulation and its implementation for the antiproton–hydrogen collision system. The computed EDCS and verification of the zero-overlap condition are presented and discussed in Section~\ref{sec:results}. Finally, the main findings and conclusions are summarized in Section~\ref{sec:conclusions}.

Atomic units, characterized by $\hbar=m_e=e=4\pi\epsilon_0=1$, are used unless otherwise stated.

\section{General theory}
\label{sec:theory}

Within the framework of the semi-classical impact parameter method, the antiproton's motion is represented by a straight-line trajectory $R(t,b)=(b,0,vt)$, where \(b\) is the impact parameter and \(z = vt\). The TDSE to be solved is

\begin{equation}
    i \frac{d}{dt} \ket{\psi(t)} = \hat{H}(t) \ket{\psi(t)} ,
\label{eq:tdse}
\end{equation}
and, in a target-centered reference frame, the Hamiltonian in position space takes the form

\begin{equation}
	\hat H(t) = - \frac{1}{2}\nabla^2 - \frac{1}{r} + \frac{1}{|{\bf r}- {\bf R}(t)|},
\label{eq:hcoord}
\end{equation} 
and asymptotically approaches $\hat H_0  = - \frac{1}{2}\nabla^2 - \frac{1}{r}$ for $t\rightarrow \pm \infty$.

In the coupled-channel approach, solving the TDSE necessitates expressing the system in an $N$-dimensional basis and determining the time evolution of a projected state vector via

\begin{equation}
	i \frac{d}{dt} \ket{\psi_P (t)} = \hat P \hat H(t) \hat P \ket{\psi_P (t)} 
\label{eq:tdsep}
\end{equation} 
where $\hat P = \hat P^2$ is the projection operation onto the  $N$-dimensional subspace $ \cal P$ and the projected state vector satisfies $\ket{\psi_P (t)} =\hat P \ket{\psi_P (t)}$.

Assuming that at large times $t_f$, the Hamiltonian can be approximated by the asymptotic Hamiltonian $\hat H_0$, the projected state vector can be projected onto the eigenstates $\ket{\phi_k}$ of $\hat H_0$ to extract the transition amplitudes $a_k^P(t) = \braket{\phi_k}{\psi_P (t)}$. If the projected state vector $\ket{\psi_P (t)}$ is a good approximation to $\ket{\psi (t)}$ of Eq.~(\ref{eq:tdse}), the transition probabilities $|a_k^P(t)|^2$ are expected to be reasonably close to the exact values $p_k(t)=|\braket{\phi_k}{\psi (t)}|^2$ which are constant at $t \geq t_f$ under the condition stated above.

Asymptotic stability of the transition probabilities $|a_k^P(t)|^2$ however cannot be guaranteed, as can be seen by inspecting $i \dot a_k^P(t) = \braket{\phi_k}{i\frac{d}{dt}\psi_P (t)} $ using Eq.~(\ref{eq:tdsep}):

\begin{equation}
i \dot a_k^P(t) = \bra{\phi_k} \hat P\hat H_0 \hat P \ket{\psi_P (t)} + 
		  \bra{\phi_k} \hat P\hat V(t) \hat P \ket{\psi_P (t)},
\label{eq:ak2}
\end{equation} 
where $ \hat{V}(t) = \frac{1}{|{\bf r}- {\bf R}(t)|}$. The first term represents evolution under the unperturbed projected Hamiltonian, while the second term accounts for the time-dependent perturbation induced by the projectile’s motion. For $t \ge t_f$ the time-dependent potential does not vanish completely; if approximated by its monopole expansion term one finds

\begin{equation}
i \dot a_k^P(t) = \bra{\phi_k} \hat P\hat H_0 \hat P \ket{\psi_P (t)} + 
		  \frac{1}{R(t)} a_k^P(t)  \hskip20pt (t\ge t_f) .
\label{eq:ak3}
\end{equation}

The projected Hamiltonian can be represented in terms of its normalized eigenstates $\ket{\varphi_\ell}$ and corresponding eigenvalues $\epsilon_\ell$ as

\begin{equation}
    \hat P\hat H_0 \hat P =  \sum_{\ell=1}^{N} \ket{\varphi_\ell} \epsilon_\ell \bra{\varphi_\ell}
    \label{eq:php}
\end{equation} which, when substituted in Eq.~(\ref{eq:ak3}) yields

\begin{equation}
	i \dot a_k^P(t) = \sum_{\ell =1}^N \epsilon_\ell \braket{\phi_k}{\varphi_\ell}\braket{\varphi_\ell}{\psi_P(t)}
		  +   \frac{1}{R(t)} a_k^P(t)  \hskip20pt (t\ge t_f) .
\label{eq:ak4}
\end{equation}

Channel couplings in the first term on the right-hand side prevent asymptotic stability of the transition probabilities. However, if the condition

\begin{equation}
   \braket{\phi_k}{\varphi_\ell} = \delta_{\ell \ell'} \braket{\phi_k}{\varphi_{\ell'}} 
   \label{eq:zero_o}
\end{equation}

 for $\ell = 1,..,N$ and for a specific index $\ell'$, referred to as the zero-overlap condition, is satisfied, Eq.~(\ref{eq:ak4}) can be further reduced to 

\begin{equation}
	i \dot a_k^P(t) = \Big(\epsilon_{\ell'} +  \frac{1}{R(t)}\Big) a_k^P(t)  \hskip20pt (t\ge t_f) 
\label{eq:ak5}
\end{equation}and can be integrated to yield a constant probability
$p_k^P \equiv |a_k^P(t)|^2 $ at $t\ge t_f$.
The zero-overlap condition seems contrived but, as it turns out, can be met by suitably constructed $\cal P$ spaces.

To shed light on this issue, we consider the 
asymptotic $\hat H_0$ problem. Introducing the orthogonal projection operator $\hat Q = \hat 1 - \hat P$ with
$\hat Q \hat P = \hat P \hat Q = 0$, the Schr\"odinger equation 
$\hat H_0 \ket{\phi_k}=E_k \ket{\phi_k}$ can be decomposed into \cite{Feshbach1962}

\begin{eqnarray}
	\hat P (\hat H_0 - E_k )\hat P \ket{\phi_k} &=& -\hat P \hat H_0 \hat Q \ket{\phi_k}  ,
	\label{eq:pse} \\
        \hat Q (\hat H_0 - E_k )\hat Q \ket{\phi_k} &=& -\hat Q \hat H_0 \hat P \ket{\phi_k}  .
	\label{eq:qse}
\end{eqnarray}
Let us assume for a moment that the $\cal P$ space is chosen one-dimensional such that Eq.~(\ref{eq:php})
has only one term, say for $\ell=\ell'$. In this case, the $\cal P$-space equation~(\ref{eq:pse}) simplifies
to
\begin{equation}
	\hat P (\epsilon_{\ell'} - E_k) \hat P \ket{\phi_k} = -\hat P \hat H_0 \hat Q \ket{\phi_k}.
\end{equation}
If the eigenvalues match ($\epsilon_{\ell'}= E_k$), the equation decouples from the 
$\cal Q$-space equation~(\ref{eq:qse}).
For an $N$-dimensional $\cal P$ space, $\hat P \hat H_0 \hat Q \ket{\phi_k}=0$ is achieved if
in addition to the matching condition $\epsilon_{\ell'}= E_k$ the overlaps
between the eigenstates of $\hat P \hat H_0 \hat P$ for $\ell\neq \ell'$ and the eigenstate of $\hat H_0$ vanish: 
\begin{equation}
	\sum_{\ell \neq \ell'} \ket{\varphi_\ell} (\epsilon_\ell - E_k) \braket{\varphi_\ell}{\phi_k} = 0 \;\;
         \Leftrightarrow  \;\; \braket{\varphi_\ell}{\phi_k} =0 \; \forall \ell \neq \ell'.
\end{equation}
This is the zero-overlap condition (\ref{eq:zero_o}) invoked to turn Eq.~(\ref{eq:ak4}) 
into Eq.~(\ref{eq:ak5})
to ensure asymptotic stability of the transition probabilities $p_k^P$.
We are interested in transitions to the continuum in this work. This implies that an $N$-dimensional
$\cal P$ space spanned by a set of linearly-independent Hilbert space vectors will yield a set of
eigenvalues $\{\epsilon_{\ell}\}$ upon diagonalization of the projected Hamiltonian $\hat P \hat H_0 \hat P$,
whose non-negative elements match a set of continuum eigenvalues $\{E_k\}$ of $\hat H_0$. 
It is not at all obvious that the zero-overlap condition can be met for any or all eigenstates {$\phi_k$} corresponding to {$E_k$}. However, for the Coulomb
problem and a $\cal P$ space spanned by a Laguerre basis 
McGovern {\it et al.} numerically demonstrated that this is indeed the case~\cite{McGovern2009}. Abdurakhmanov \textit{et al.} subsequently provided a proof for this finding~\cite{Abdurakhmanov2011a}.
A more general analysis of the zero-overlap condition including an alternative proof for the Laguerre basis will be presented elsewhere \cite{Tsui2026}.
In the next section, we demonstrate that OC-BGM basis sets constructed to solve the $\cal P$-space 
TDSE Eq.~(\ref{eq:tdsep}) for the antiproton--hydrogen collision problem meet the condition in good approximation.
This implies that projecting the OC-BGM
solutions onto Coulomb waves at the matching energy values yields
asymptotically stable transition probabilities for electron emission and opens the door to
differential cross-section calculations.

\section{OC-BGM for antiproton--hydrogen collisions}
\label{sec:ocbgm}

The present implementation builds upon earlier applications of the BGM to ion–atom and antiproton–atom collisions, which were restricted to total ionization calculations where the method was shown to efficiently represent both bound and continuum electronic dynamics within a finite pseudostate basis \cite{Kirchner1997, Elizaga1999, Keim2003}. The present OC-BGM formulation is the first to extract differential cross-sections for the single-electron antiproton–hydrogen system.

A stationary basis is constructed to transform the $\cal P$-space TDSE Eq.~(\ref{eq:tdsep}) into a 
set of coupled-channel equations for the expansion coefficients, which are subsequently solved numerically. The projected state vector is expressed as

\begin{equation}
	\ket{\psi_P (t)} = \sum_{j=1}^{j_{\rm max}}\sum_{J=0}^{J_{\rm max}} c_{jJ}(t) \ket{jJ}
    \label{pseudo}
\end{equation}
with
\begin{equation}
	\ket{jJ} =  W_t^J \ket{j0} ,
    \label{eq:Pseudostate}
\end{equation}
in which $\ket{j0}$ are bound-state solutions of the $\hat H_0$ eigenvalue problem, i.e., hydrogenic
atomic orbitals (AOs), and $W_t$ is the Yukawa-regularized target potential 
\begin{equation}
	W_t(r) = \frac{1}{r} \big(1-e^{-r}\big) .
\label{eq:wt}
\end{equation}
Resolving the AO index $j$ into the quantum numbers $nlm$, the
propagated wave function $\psi_P ({\bf r},t) = \braket{{\bf r}}{\psi_P(t)}$ can be written more explicitly in
terms of an angular momentum decomposition as
\begin{equation}
	\psi_P ({\bf r},t) = \sum_{lm} P_{lm} (r,t) Y_{lm} (\Omega_r)
\end{equation}
with spherical harmonics $Y_{lm}(\Omega_r)$ in position space and (complex-valued) time-dependent radial functions
\begin{equation}
	P_{lm} (r,t) = \sum_{n=1}^{n_{\rm max}} \sum_{J=0}^{J_{\rm max}}c_{nlm,J} (t) W_t^J(r) R_{nl}(r) ,
\end{equation}
which consist of combinations of time-dependent expansion coefficients, powers of the 
regularized potential Eq.~(\ref{eq:wt}), and 
radial wave functions of the Schr\"odinger hydrogen problem.

The ionization transition amplitudes are determined by projecting the $\cal P$-space wavefunction at time $t_f$ onto Coulomb wavefunctions

\begin{equation}
\phi_{{\textbf{k}}} (\textbf{r}) = 4\pi\sum_{LM}i^Le^{i\sigma_L}\frac{w_L(r)}{kr}Y^*_{LM}(\Omega_k)Y_{LM}(\Omega_r),
\label{CW}
\end{equation}
which constitute the positive-energy eigenstates of the Hamiltonian $\hat{H_0}$. Here $w_L(r)$ is the radial solution of the Coulomb wave equation, which involves the confluent hypergeometric function, $\sigma_L = arg(\Gamma(L+1+i\eta))$ is the Coulomb phase shift, and $\eta = -\frac{1}{k}$ is the Sommerfeld parameter. The solid angle {$(\Omega_k)$} refers to the wave vector \textbf{k}. Utilizing the orthonormality of the spherical harmonics in position space simplifies the summation, and the resulting expression takes the form

\begin{equation}
      \braket{\phi_{{\textbf{k}}}}
    {\psi_P(t_f)} = \sum_{lm} Q_{lm}(k,t_f) Y_{lm}(\Omega_k) 
    \label{eq:wfproj}
\end{equation}
with
\begin{equation}
 Q_{lm}(k,t_f)  =  \int_0^\infty dr r^2  4\pi 
    (-i)^le^{-i\sigma_l}\frac{w_l(r)}{kr} P_{lm} (r,t_f).
\end{equation}

Using the orthonormality relationship of the spherical harmonics once again, the total ionization probability is

\begin{equation}
     P_{ion} = \int d^3k |\braket{\phi_{{\textbf{k}}}}{\psi_P(t_f)}|^2  = \sum_{lm} \int_0^\infty | Q_{lm}(k,t_f)|^2 k^2 dk,
\end{equation}
which serves as the basis for computing the total cross section (TCS). The TCS is obtained by integrating the ionization probability over all possible impact parameters b, expressed in cylindrical coordinates as

\begin{equation}
    \sigma_{Total} = 2 \pi \int^\infty_0 P_{ion} b db.
\end{equation}

The EDCS can be readily obtained by differentiating with respect to the electron emission energy E, and using the relation $k = \sqrt{2E}$ 
\begin{equation}
    \frac{d\sigma}{dE} = \frac{1}{k} \frac{d\sigma}{dk} = 2 \pi \sum_{lm} \int^\infty_0b|Q_{lm}(k,b)|^2kdb.
\end{equation}

An alternative approach for computing the EDCS (and TCS) is detailed in Ref.\cite{Abdurakhmanov2011b}. In this method, the EDCS at discrete eigenenergies corresponding to the spectral decomposition introduced in Eq.(\ref{eq:php}) is identified as the partial cross section for population of the eigenstate $\ket{\varphi_\ell}$ divided by a weight factor which for a given angular momentum $l$ is defined as 

\begin{equation}
    w^l_\ell = \frac{\epsilon_{\ell+1,l} - \epsilon_{\ell-1,l}}{2}.
\end{equation}

In the following section \ref{sec:results}, we compare the results of the EDCS obtained using this approach with those from the projection method, highlighting the necessity of enforcing the zero-overlap condition to ensure physically meaningful and numerically stable results.

\section{Results}
\label{sec:results}

Before presenting results, it is useful to provide some contextual remarks regarding the basis construction and numerical parameters employed in this work. 
The basis is constructed using the pseudostate expansion of Eq.~\ref{pseudo}, with a maximum hierarchy index of $J_{\text{max}} = 9$ and a highest included orbital angular momentum of $l_{\text{max}} = 3$.
Table~\ref{tab:jmax} summarizes the maximum hierarchy for each $(n,l)$, and the resulting $\mathcal{P}$-space is spanned by 113 linearly independent basis states.
Table~\ref{tab:eigenvalues} lists for each $l$ the wave numbers corresponding to the first five positive eigenvalues obtained by diagonalizing $\hat{H}_0$ in the basis.
The time-dependent coefficients are obtained by numerically propagating the coupled-channel equations from $z= - 80$ to $z_f = v t_f = 80$ a.u., for antiproton impact energies ranging from 10 keV to 200 keV.

\begin{table*}[h!]
\centering
\begin{minipage}{0.45\linewidth}
\centering
\begin{tabular}{c|cccc}
\toprule
Orbital ($l$) & $n=1$ & $n=2$ & $n=3$ & $n=4$ \\
\midrule
$s$ ($l=0$) & 0 & 1 & 2 & 4 \\
$p$ ($l=1$) & -- & 2 & 3 & 5 \\
$d$ ($l=2$) & -- & -- & 5 & 5 \\
$f$ ($l=3$) & -- & -- & -- & 9 \\
\bottomrule
\end{tabular}
\caption{Maximum hierarchy $J_{\max}$ for each orbital and principal quantum number $n$ in the 113-state basis consisting of the 20 bound nl|m| eigenstates of hydrogen for $n\leq 4$ and 93 pseudostates.}
\label{tab:jmax}
\end{minipage}
\hfill
\begin{minipage}{0.45\linewidth}
\centering
\begin{tabular}{c|ccccc}
\toprule
Orbital ($l$) & $k_1$ & $k_2$ & $k_3$ & $k_4$ & $k_5$ \\
\midrule
$s$ ($l=0$) & 0.04 & 0.35 & 0.76 & 1.49 & 2.93 \\
$p$ ($l=1$) & 0.21 & 0.43 & 0.72 & 1.13 & 1.73 \\
$d$ ($l=2$) & 0.21 & 0.44 & 0.73 & 1.12 & 1.69 \\
$f$ ($l=3$) & 0.11 & 0.41 & 0.79 & 1.33 & 2.13 \\
\bottomrule
\end{tabular}
\caption{Wave numbers (in atomic units) corresponding to positive-energy eigenvalues of the projected Hamiltonian $\hat P\hat H_0 \hat P$ for $l = 0-3$ via $k_\ell$ = $\sqrt{2 \epsilon_\ell}$.}
\label{tab:eigenvalues}
\end{minipage}
\end{table*}

To assess whether the basis satisfies the zero-overlap condition, we examine the projection of each $\hat P\hat H_0 \hat P$ eigenstate onto Coulomb continuum waves and identify the wave numbers at which the overlaps exhibit minima.
McGovern \textit{et al.}~\cite{McGovern2009} studied a similar effect for their Laguerre-type pseudostates by evaluating what they termed \textit{distribution functions}, defined from the projection onto Coulomb waves and expressed as a function of the continuum momentum~$k$:

\begin{equation}
    f_{\ell l}(k) = k^2 \int d\Omega_k |\braket{\phi_{\textbf{k}}}{\psi_{\ell lm}}|^2,
\end{equation}
such that
\begin{equation}
    \int d^3k |\braket{\phi_{\textbf{k}}}{\psi_{\ell lm}}|^2 = \int_0^\infty f_{\ell l}(k) dk.
\end{equation}

Here $\psi_{\ell lm}$ represents McGovern's Laguerre based pseudostates. As mentioned in Sec \ref{sec:theory}, they observed that each distribution function vanishes at the wave numbers \(k_{\ell,l}\) which correspond to the eigenenergies \( \epsilon_{\ell,l} \) of the other pseudostates through
\begin{equation}
   \epsilon_{\ell,l} = \frac{k_{\ell,l}^2}{2}.
   \label{eq:eigenenergy}
\end{equation}
This behaviour is the \textit{zero-overlap condition}.

\begin{figure*}[h]
    \centering
    \begin{subfigure}[b]{0.49\linewidth}
        \centering
        \includegraphics[width=3.2in,height=2in]{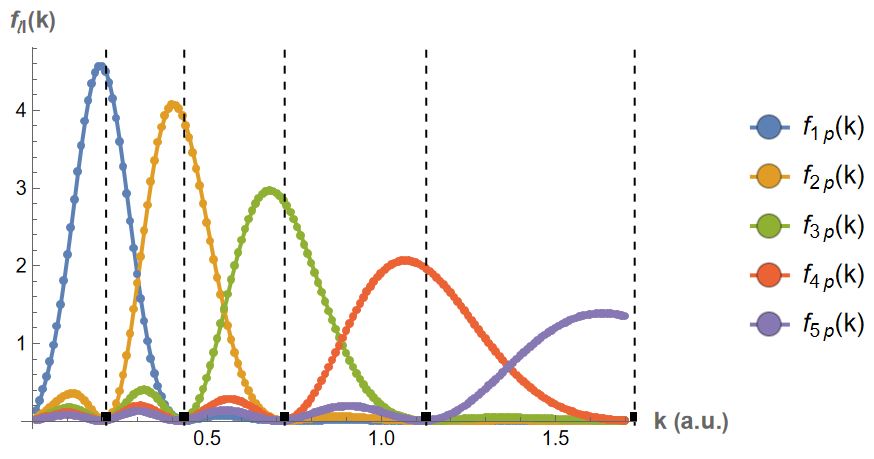}
        \caption{}
        \label{Fig.wt113l1}
    \end{subfigure}%
    ~ 
    \begin{subfigure}[b]{0.4\linewidth}
        \centering
        \includegraphics[width=3in,height=2in]{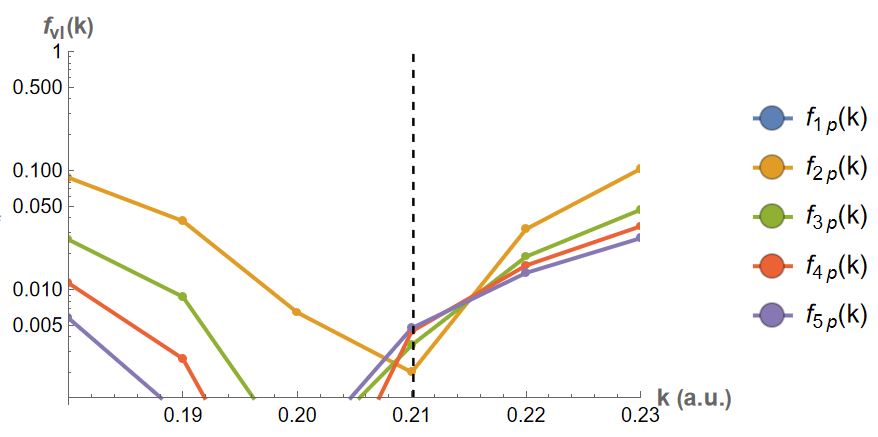}
        \caption{}
        \label{Fig.wt113l1Log}
    \end{subfigure}
    \caption{ Distribution functions (see text) obtained from OC-BGM calculations for $l=1$. The black squares in Panel.~\ref{Fig.wt113l1} are the wave numbers for $l = 1$ in Table \ref{tab:eigenvalues}. Panel.~\ref{Fig.wt113l1Log} shows the functions on a logarithmic scale in the vicinity of the first positive eigenvalue.}
    \label{Fig:zero_o}
\end{figure*}

Figures~\ref{Fig.wt113l1} and \ref{Fig.wt113l1Log} show OC-BGM based distribution functions for the first five positive-energy \textit{p}-orbital eigenstates of the projected Hamiltonian in our 113-state basis.
Although these functions exhibit pronounced minima near the pseudostate eigenenergies, they do not reach exact zeros, indicating that the zero-overlap condition is satisfied only approximately in the present calculation.

\begin{figure}[h]
    \centering
    \includegraphics[width=0.5\linewidth]{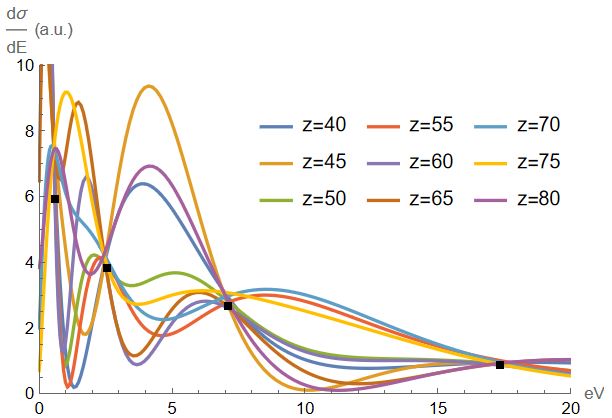}
    \caption{EDCS for $l = 1$ calculated at $z_f$ $\in$ $[40,80]$ for an impact energy of 30 keV plotted as a function of the electron emission energy. The black squares are the EDCS obtained from the summation method \cite{Abdurakhmanov2011b}.}
    \label{fig:edcsl1}
\end{figure}

Figure~\ref{fig:edcsl1} highlights the significance of the zero-overlap condition. It presents the EDCS for $l=1$ as a function of electron ejection energy and shows that the results obtained from carrying out the projections (\ref{eq:wfproj}) at different values of \( z_f \in [40,80] \) closely agree with the EDCS computed using Abdurakhmanov \textit{et al.'s} summation method \cite{Abdurakhmanov2011b} at the pseudostate eigenenergies, except at the lowest eigenvalue. In between these discrete points, the projections yield distance-dependent results as a consequence of the channel couplings in Eq. (\ref{eq:ak4}).
When plotted together, the EDCS curves intersect forming "nodes" around the eigenvalues of the projected Hamiltonian.

These behaviors as shown in Fig.\ref{Fig:zero_o} and Fig.\ref{fig:edcsl1} for $l = 1$ have been tested for different basis sizes and other orbital angular momenta, and in each case, similar structures have been found.
It is important to note that the values of the EDCS between the eigenvalues \( \epsilon_\ell \) cannot be reliably determined, as the transition amplitudes are unstable in those regions.
This emphasizes the necessity of the zero-overlap condition and that physically meaningful results can only be obtained at the eigenvalues. 

In principle, these structures can be flattened by increasing the size of the basis set thereby increasing the number of positive pseudostate eigenvalues.
In practice, however, this is not easily achieved due to numerical limitations.
As the basis size increases, the system approaches saturation, and the eigenstates of the projected Hamiltonian begin to lose linear independence, thereby compromising numerical accuracy and stability. This feature is a detriment of the BGM basis set construction scheme using highly non-orthonormal states (\ref{eq:Pseudostate}).

\begin{figure}[h]
    \centering
    \includegraphics[width=0.5\linewidth]{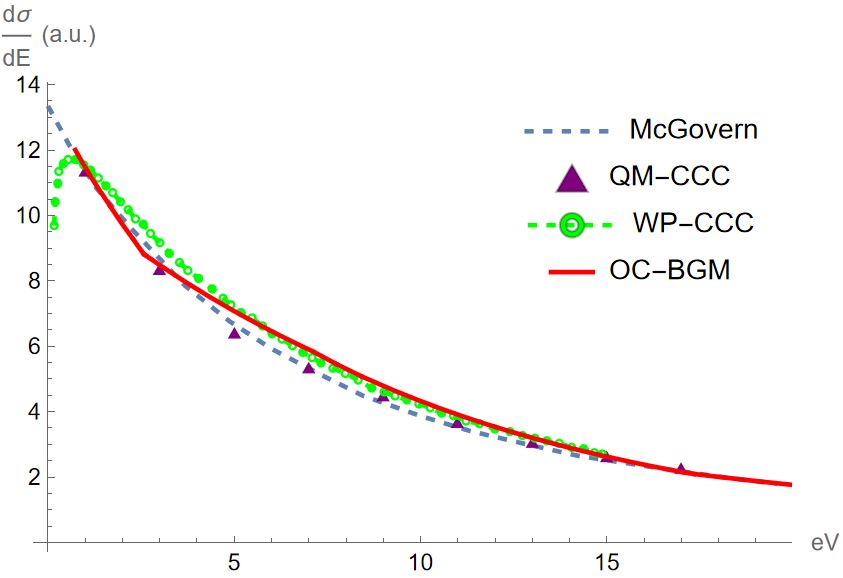}
    \caption{EDCS obtained from the present calculations at 30 keV compared with results from McGovern's pseudostate approaches \cite{McGovern2010}, QM-CCC ~\cite{Abdurakhmanov2011b}, and WP-CCC \cite{Abdurakhmanov2016}.
}
    \label{fig:EDCS}
\end{figure}

To obtain the "total" EDCS, we first average the EDCS for each orbital angular momentum $l$ across \( z_f =[40,80] \) at the eigenvalues. Then we construct exponential piecewise functions to interpolate the EDCS in energy regions not directly accessible due to the invalidity of the zero-overlap condition. The exponential form is chosen to reflect the expected smooth decay of the EDCS. This is illustrated in Appendix \ref{appendix:PartialWave}.
The resulting EDCS for each $l$ are subsequently summed to obtain the total EDCS. Figure ~\ref{fig:EDCS} compares our results at 30 keV impact energy to McGovern's pseudostate-based approach~\cite{McGovern2010}, the quantum mechanical convergent close-coupling (QM-CCC) approach~\cite{Abdurakhmanov2011b}, and the wave packet CCC method (WP-CCC)~\cite{Abdurakhmanov2016}.
We observe that the interpolated OC-BGM results closely track the benchmark QM-CCC and WP-CCC calculations across the full energy range shown.

\begin{figure}[h!]
    \centering
    \includegraphics[width=0.5\linewidth]{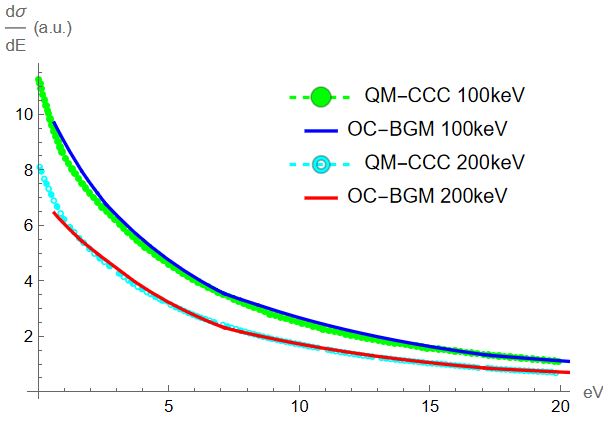}
    \caption{EDCS obtained from the present calculations at 100 and 200 keV compared with results from QM-CCC ~\cite{Abdurakhmanov2011b}.
    }
    \label{fig:EDCS100-200}
\end{figure}

Having established the reliability of the interpolated total EDCS at an intermediate impact energy, we now examine its behavior at 100 keV and 200 keV.
This is shown in Fig.\ref{fig:EDCS100-200}.
For both energies, the OC-BGM EDCS reproduces the QM-CCC results across the full electron energy range shown, demonstrating good agreement between the two approaches.

\begin{figure}[h]
    \centering
    \includegraphics[width=0.5\linewidth]{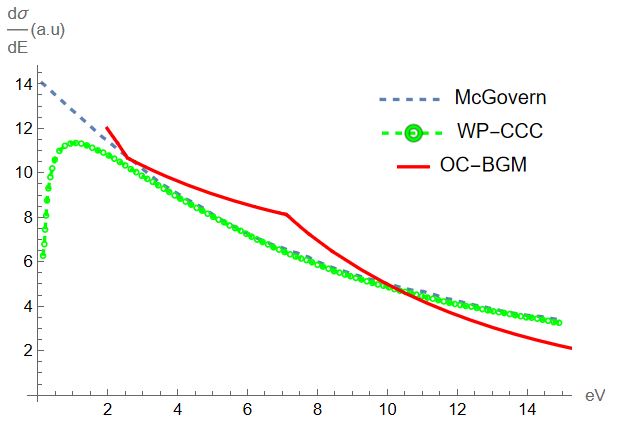}
    \caption{EDCS obtained from the present calculations at 10 keV compared with results from McGovern \cite{McGovern2010}, and QM-CCC \cite{Abdurakhmanov2011b} pseudostate approaches.
    }
    \label{fig:EDCS10}
\end{figure}

To further assess the range of validity of the present approach, we also examine the EDCS at a lower antiproton impact energy of 10 keV in Fig.~\ref{fig:EDCS10}.
In this low-energy regime, the OC-BGM results exhibit un-physical structures and noticeable deviations from the WP-CCC and 
McGovern \textit{et al.} pseudostate calculations across the electron energy range shown. The details provided in Appendix \ref{appendix:PartialWave} indicate that the $l = 1$ contribution induces most of the structure seen in Fig. \ref{fig:EDCS10}. We were not able to identify a simple reason for this behavior and conclude that the present OC-BGM formulation becomes less accurate for EDCS calculations in the low-impact-energy regime.

\section{Conclusions}
\label{sec:conclusions}

In this work, we investigated electron emission from hydrogen atoms induced by antiproton impact using the OC-BGM within a semi-classical impact-parameter framework. 
The reduction of the full three-body problem to a single-centre formulation enabled  a simpler treatment of the time-dependent Schrödinger equation, where BGM aims to find accurate results with high efficiency using a relatively low number of pseudostates that are constructed out of a regularized Yukawa potential. The present study is the first application of BGM to a differential cross section calculation.

A central focus of the analysis was the role of the zero-overlap condition in ensuring the numerical stability and physical reliability of the computed EDCS. 
By projecting the pseudostate-evolved wavefunction onto Coulomb continuum states, we demonstrated that the zero-overlap condition is approximately satisfied at the eigenvalues of the projected unperturbed Hamiltonian. 
When this condition holds, the extracted EDCS values remain stable across different projection distances, underscoring its importance for obtaining physically meaningful observables.

Away from the pseudostate eigenenergies, the EDCS exhibits fluctuations due to incomplete convergence, and physically reliable values cannot be obtained.
To obtain smooth cross-section profiles suitable for comparison with other theoretical approaches (such as QM-CCC, WP-CCC, and McGovern’s coupled-pseudostate method), exponential interpolation was employed between the eigenvalue points. The total EDCS, reconstructed by summing the interpolated partial-wave contributions, shows good agreement with results from these methods except at the lowest impact energy investigated. 
These results validate the OC-BGM as a compact and physically transparent framework for modeling differential ionization in antiproton-hydrogen collisions at intermediate energies.

\section*{Acknowledgments}
We thank Janakan Sivasubramanium and Marko Horbatsch for insightful discussions and helpful input related to the zero-overlap condition. Financial support from the Natural Sciences and Engineering Research Council of Canada (Grant No. RGPIN-2025-06277) is gratefully acknowledged.

\newpage

\bibliography{Antiproton_EDCS}

@article{Feshbach1962,
title = {A unified theory of nuclear reactions. II},
journal = {Annals of Physics},
volume = {19},
number = {2},
pages = {287-313},
year = {1962},
issn = {0003-4916},
doi = {https://doi.org/10.1016/0003-4916(62)90221-X},
url = {https://www.sciencedirect.com/science/article/pii/000349166290221X},
author = {Herman Feshbach},
}

@article{Ludde1996,
  author = {H. J. Lüdde and A. Henne and T. Kirchner and M. Dreizler},
  title = {Optimized dynamical representation of the solution of time-dependent quantum mechanical few-body problems},
  journal = {J. Phys. B: At. Mol. Opt. Phys.},
  volume = {29},
  pages = {4423--4444},
  year = {1996},
  doi = {10.1088/0953-4075/29/19/015}
}

@article{Kroneisen1999,
  author = {O. J. Kroneisen and H. J. Lüdde and T. Kirchner and R. M. Dreizler},
  title = {The basis generator method: optimized dynamical representation of the solution of time-dependent quantum problems},
  journal = {J. Phys. A: Math. Gen.},
  volume = {32},
  pages = {2141--2156},
  year = {1999},
  doi = {10.1088/0305-4470/32/11/009}
}

@article{Zapukhlyak2005,
  author = {M. Zapukhlyak and T. Kirchner and H. J. L\"udde and S. Knoop and R. Morgenstern and R. Hoekstra},
  title = {Inner- and outer-shell electron dynamics in proton collisions with sodium atoms},
  journal = {J. Phys. B: At. Mol. Opt. Phys.},
  volume = {38},
  number = {14},
  pages = {2353--2369},
  year = {2005},
  doi = {10.1088/0953-4075/38/14/003}
}

@article{Kirchner1997,
  author = {Kirchner, T. and Gulyás, L. and Lüdde, H. J. and Henne, A. and Engel, E. and Dreizler, R. M.},
  title = {Electronic exchange effects in p + Ne and p + Ar collisions},
  journal = {Phys. Rev. Lett.},
  volume = {79},
  pages = {1658--1661},
  year = {1997},
  doi = {10.1103/PhysRevLett.79.1658}
}

@article{Kirchner1998,
  author = {Kirchner, T. and Gulyás, L. and Lüdde, H. J. and Engel, E. and Dreizler, R. M.},
  title = {Time-dependent density-functional calculations for antiproton collisions with hydrogen atoms},
  journal = {Phys. Rev. A},
  volume = {58},
  number = {3},
  pages = {2063--2076},
  year = {1998},
  doi = {10.1103/PhysRevA.58.2063}
}

@article{Elizaga1999,
  author = {Elizaga, D. and Errea, L. F. and Gorfinkiel, J. D. and Illescas, C. and Méndez, L. and Macías, A. and Riera, A. and Rojas, A. and Kroneisen, O. J. and Kirchner, T. and L\"udde, H. J. and Henne, A. and Dreizler, R. M.},
  title = {Theoretical analysis of electron capture and electron loss in Be$^{4+}$ + H$_2$ and H$^+$ + H$_2$ collisions},
  journal = {J. Phys. B: At. Mol. Opt. Phys.},
  volume = {32},
  number = {4},
  pages = {857--875},
  year = {1999},
  doi = {10.1088/0953-4075/32/4/005}
}

@article{Keim2003,
  author = {M. Keim and A. Achenbach and H. J. L\"udde and T. Kirchner},
  title = {Microscopic response effects in collisions of antiprotons with helium atoms and lithium ions},
  journal = {Phys. Rev. A},
  volume = {67},
  pages = {062711},
  year = {2003},
  doi = {10.1103/PhysRevA.67.062711}
}

@article{Keim2005,
  author = {Keim, M. and Achenbach, A. and L\"udde, H. J. and Kirchner, T.},
  title = {Time-dependent density functional theory calculations for collisions of bare ions with helium},
  journal = {Nucl. Instrum. Methods Phys. Res., Sect. B},
  volume = {233},
  pages = {240--243},
  year = {2005},
  doi = {10.1016/j.nimb.2005.03.114}
}

@article{Henkel2009,
  author = {Henkel, N. and Keim, M. and Lüdde, H. J. and Kirchner, T.},
  title = {Density-functional-theory investigation of antiproton-helium collisions},
  journal = {Phys. Rev. A},
  volume = {80},
  pages = {032704},
  year = {2009},
  doi = {10.1103/PhysRevA.80.032704}
}

@article{Leung2015,
  author = {Leung, A. C. K. and Kirchner, T.},
  title = {Independent-electron analysis of the x-ray spectra from single-electron capture in Ne$^{10+}$ collisions with He, Ne, and Ar atoms},
  journal = {Phys. Rev. A},
  volume = {92},
  pages = {032712},
  year = {2015},
  doi = {10.1103/PhysRevA.92.032712}
}

@article{Ludde2018,
  author = {Lüdde, H. J. and Horbatsch, M. and Kirchner, T.},
  title = {A screened independent atom model for the description of ion collisions from atomic and molecular clusters},
  journal = {Eur. Phys. J. B},
  volume = {91},
  pages = {99},
  year = {2018},
  doi = {10.1140/epjb/e2018-90165-x}
}

@article{Leung2019,
  author = {Leung, A. C. K. and Kirchner, T.},
  title = {Proton impact on ground and excited states of atomic hydrogen},
  journal = {Eur. Phys. J. D},
  volume = {73},
  number = {246},
  year = {2019},
  doi = {10.1140/epjd/e2019-100380-x}
}

@article{Ludde2021,
  author = {Lüdde, H. J. and Horbatsch, M. and Kirchner, T.},
  title = {Calculation of energy loss in antiproton collisions with many-electron systems using Ehrenfest’s theorem},
  journal = {Phys. Rev. A},
  volume = {104},
  number = {3},
  pages = {032813},
  year = {2021},
  doi = {10.1103/PhysRevA.104.032813}
}

@article{Leung2022,
  author = {A. C. K. Leung and T. Kirchner},
  title = {Two-center basis generator method calculations for Li$^{3+}$, C$^{3+}$, and O$^{3+}$ ion impact on ground-state hydrogen},
  journal = {Atoms},
  volume = {10},
  number = {1},
  pages = {11},
  year = {2022},
  doi = {10.3390/atoms10010011}
}

@article{Schultz1989,
  author = {Schultz, D. R.},
  title = {Comparison of single-electron removal processes in collisions of electrons, positrons, protons, and antiprotons with hydrogen and helium},
  journal = {Phys. Rev. A},
  volume = {40},
  number = {5},
  pages = {2330--2336},
  year = {1989},
  doi = {10.1103/PhysRevA.40.2330}
}

@article{Schultz1996,
  author = {Schultz, D. R. and Krstić, P. S. and Reinhold, C. O. and Wells, J. C.},
  title = {Ionization of hydrogen and hydrogenic ions by antiprotons},
  journal = {Phys. Rev. Lett.},
  volume = {76},
  number = {16},
  pages = {2882--2885},
  year = {1996},
  doi = {10.1103/PhysRevLett.76.2882}
}

@article{Sidky1998,
  author = {E. Y. Sidky and C. D. Lin},
  title = {Quantum Mechanical Calculation of Ejected Electron Spectra for Ion–Atom Collisions},
  journal = {J. Phys. B: At. Mol. Opt. Phys.},
  volume = {31},
  pages = {2949--2960},
  year = {1998},
doi = {10.1088/0953-4075/31/13/013}
}

@article{Pons1999,
  author = {Pons, B.},
  title = {Monocentric close-coupling expansion to provide ejected electron distributions for ionization in atomic collisions},
  journal = {Phys. Rev. Lett.},
  volume = {84},
  number = {20},
  pages = {4569--4572},
  year = {1999},
  doi = {10.1103/PhysRevLett.84.4569}
}

@article{Pons2000,
  author = {Pons, B.},
  title = {Ability of monocentric close-coupling expansions to describe ionization in atomic collisions},
  journal = {Phys. Rev. A},
  volume = {63},
  number = {1},
  pages = {012704},
  year = {2000},
  doi = {10.1103/PhysRevA.63.012704}
}

@article{Igarashi2000,
  author = {A. Igarashi and S. Nakazaki and A. Ohsaki},
  title = {Ionization of atomic hydrogen by antiproton impact},
  journal = {Phys. Rev. A},
  volume = {61},
  pages = {062712},
  year = {2000},
  doi = {10.1103/PhysRevA.61.062712}
}

@article{Sakimoto2000,
  author = {K. Sakimoto},
  title = {Excitation and ionization in p̄ + H collisions calculated by direct numerical solution using Laguerre meshes},
  journal = {J. Phys. B: At. Mol. Opt. Phys.},
  volume = {33},
  pages = {5165--5175},
  year = {2000},
  doi = {10.1088/0953-4075/33/22/317}
}

@article{Azuma2001,
  author = {J. Azuma and N. Toshima and K. Hino and A. Igarashi},
  title = {B-spline expansion of scattering equations for ionization of atomic hydrogen by antiproton impact},
  journal = {Phys. Rev. A},
  volume = {64},
  pages = {062704},
  year = {2001},
  doi = {10.1103/PhysRevA.64.062704}
}

@article{Tong2001,
  author = {Tong, X.-M. and Watanabe, T. and Kato, D. and Ohtani, S.},
  title = {Ionization of atomic hydrogen by antiproton impact: A direct solution of the time-dependent Schrödinger equation},
  journal = {Phys. Rev. A},
  volume = {64},
  number = {2},
  pages = {022711},
  year = {2001},
  doi = {10.1103/PhysRevA.64.022711}
}

@article{Toshima2001,
  author = {Toshima, N.},
  title = {Two- and one-center close-coupling calculations for ionization of atomic hydrogen by antiproton impact},
  journal = {Phys. Rev. A},
  volume = {64},
  number = {2},
  pages = {024701},
  year = {2001},
  doi = {10.1103/PhysRevA.64.024701}
}

@article{Jones2002,
  author = {Jones, S. and Madison, D. H.},
  title = {Scaling behavior of the fully differential cross section for ionization of hydrogen atoms by the impact of fast elementary charged particles},
  journal = {Phys. Rev. A},
  volume = {65},
  number = {5},
  pages = {052727},
  year = {2002},
  doi = {10.1103/PhysRevA.65.052727}
}

@article{Voitkiv2003,
  author = {A. B. Voitkiv and J. Ullrich},
  title = {Three-body Coulomb dynamics in hydrogen ionization by protons and antiprotons at intermediate collision velocities},
  journal = {Phys. Rev. A},
  volume = {67},
  pages = {062703},
  year = {2003},
  doi = {10.1103/PhysRevA.67.062703}
}

@article{Sahoo2004,
  author = {S. Sahoo and S. C. Mukherjee and H. R. J. Walters},
  title = {Ionization of atomic hydrogen and He$^{+}$ by slow antiprotons},
  journal = {J. Phys. B: At. Mol. Opt. Phys.},
  volume = {37},
  pages = {3227--3237},
  year = {2004},
  doi = {10.1088/0953-4075/37/16/001}
}

@article{McGovern2009,
  author = {McGovern, M. and Assafrão, D. and Mohallem, J. R. and Whelan, C. T. and Walters, H. R. J.},
  title = {Differential and total cross sections for antiproton-impact ionization of atomic hydrogen and helium},
  journal = {Phys. Rev. A},
  volume = {79},
  pages = {042707},
  year = {2009},
  doi = {10.1103/PhysRevA.79.042707}
}

@article{McGovern2010,
  author = {M. McGovern and D. Assafr{\~a}o and J. R. Mohallem and C. T. Whelan and H. R. J. Walters},
  title = {Pseudostate methods and differential cross sections for antiproton ionization of atomic hydrogen and helium},
  journal = {Phys. Rev. A},
  volume = {81},
  pages = {032708},
  year = {2010},
  doi = {10.1103/PhysRevA.81.032708}
}

@article{Abdurakhmanov2011a,
  author = {I. B. Abdurakhmanov and A. S. Kadyrov and I. Bray and A. T. Stelbovics},
  title = {Coupled-channel integral-equation approach to antiproton--hydrogen collisions},
  journal = {J. Phys. B: At. Mol. Opt. Phys.},
  volume = {44},
  pages = {075204},
  year = {2011},
  doi = {10.1088/0953-4075/44/7/075204}
}

@article{Abdurakhmanov2011b,
  author = {I. B. Abdurakhmanov and A. S. Kadyrov and I. Bray and A. T. Stelbovics},
  title = {Differential ionization in antiproton--hydrogen collisions within the convergent-close-coupling approach},
  journal = {J. Phys. B: At. Mol. Opt. Phys.},
  volume = {44},
  pages = {165203},
  year = {2011},
  doi = {10.1088/0953-4075/44/16/165203}
}

@article{Kirchner2011,
  author = {Kirchner, T. and H. Knudsen},
  title = {Antiproton collisions with atoms and molecules},
  journal = {J. Phys. B: At. Mol. Opt. Phys.},
  volume = {44},
  number = {12},
  pages = {122001},
  year = {2011},
  doi = {10.1088/0953-4075/44/12/122001}
}

@article{Abdurakhmanov2016,
  author = {Abdurakhmanov, I. B. and Kadyrov, A. S. and Bray, I.},
  title = {Wave-packet continuum-discretization approach to ion-atom collisions: Nonrearrangement scattering},
  journal = {Phys. Rev. A},
  volume = {94},
  pages = {022703},
  year = {2016},
  doi = {10.1103/PhysRevA.94.022703}
}

@unpublished{Tsui2026   ,
  author = {J. Tsui and M. Horbatsch and T. Kirchner},
  title = {A Criterion for an Effective Discretization of a Continuous {Schrödinger} Spectrum Using a Pseudostate Basis},
  note = {Unpublished},
  year = {2026}
}

\newpage

\appendix

\section{Partial-Wave Energy Differential Cross Sections}
\label{appendix:PartialWave}

In this appendix we present the partial-wave energy-differential cross
sections obtained from the OC-BGM calculations at antiproton
impact energies of 30~keV and 10~keV. Each panel corresponds to a
specific orbital angular momentum $l$ contribution. The EDCS values
shown are extracted at the pseudostate eigenenergies of the projected
Hamiltonian $\hat{P}\hat{H}_0\hat{P}$, These partial-wave contributions form the
basis for constructing the interpolated total EDCS discussed in
Sec.~IV.

\begin{figure*}[ht]
\centering

\begin{minipage}{0.48\textwidth}
\centering
\includegraphics[width=\linewidth]{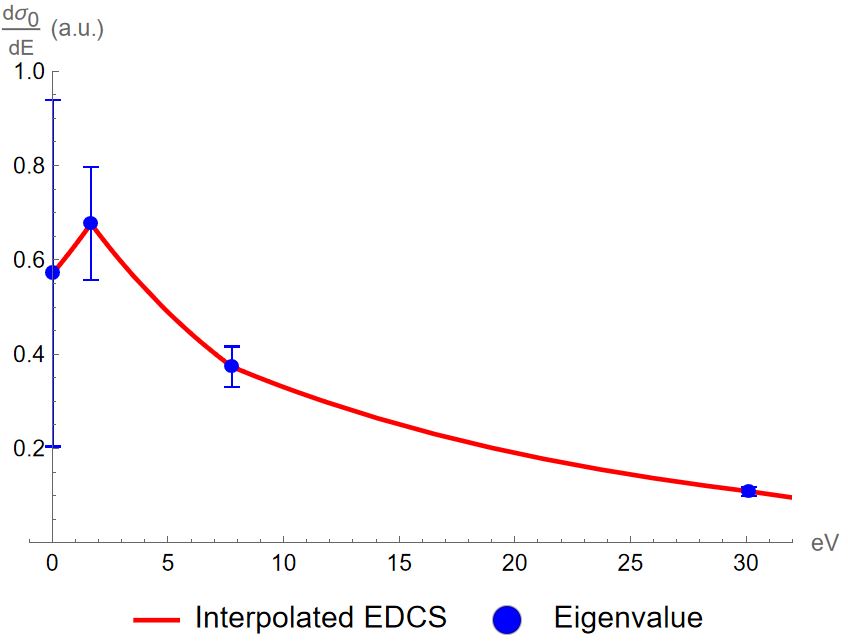}
\textbf{(a)} $l=0$
\end{minipage}
\hfill
\begin{minipage}{0.48\textwidth}
\centering
\includegraphics[width=\linewidth]{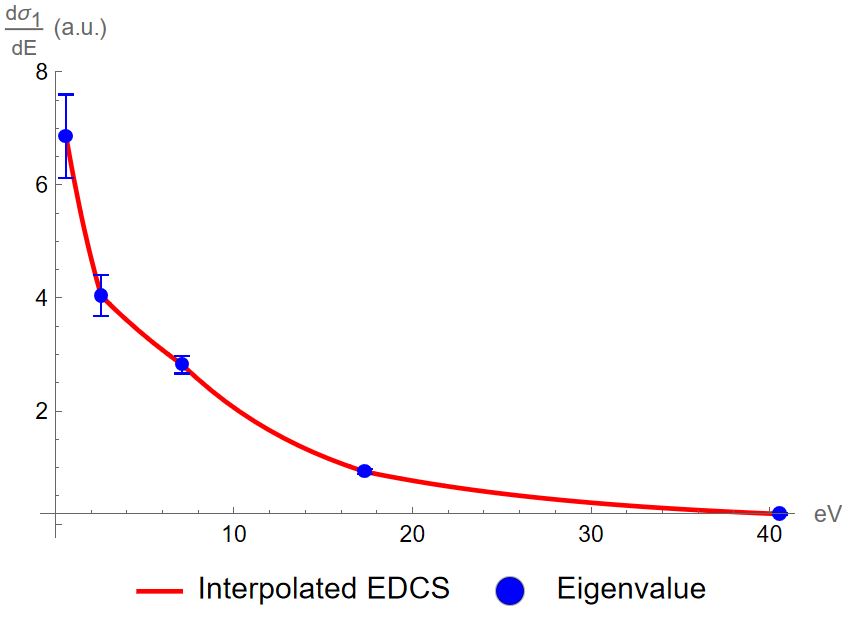}
\textbf{(b)} $l=1$
\end{minipage}

\vspace{0.5cm}

\begin{minipage}{0.48\textwidth}
\centering
\includegraphics[width=\linewidth]{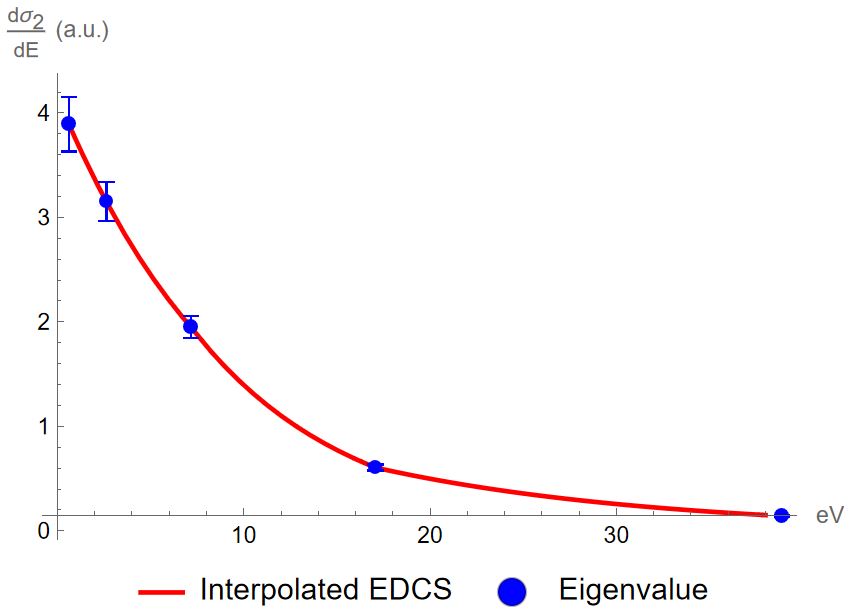}
\textbf{(c)} $l=2$
\end{minipage}
\hfill
\begin{minipage}{0.48\textwidth}
\centering
\includegraphics[width=\linewidth]{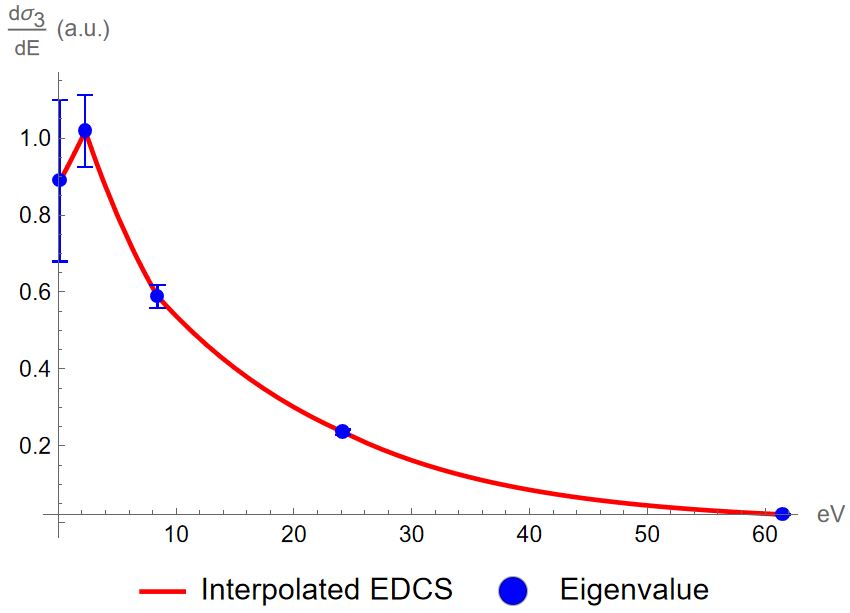}
\textbf{(d)} $l=3$
\end{minipage}

\caption{Partial-wave energy differential cross sections for
electron emission in antiproton-hydrogen collisions at an
impact energy of 30~keV calculated using the OC-BGM method.
Each panel corresponds to a different orbital angular momentum
contribution $l$. The error bars represent the standard deviations of the projections carried out at different distances in the $z_f$ $\in$ [40,80] interval.}
\label{fig.EDCS30}
\end{figure*}

\begin{figure*}[ht]
\centering

\begin{minipage}{0.48\textwidth}
\centering
\includegraphics[width=\linewidth]{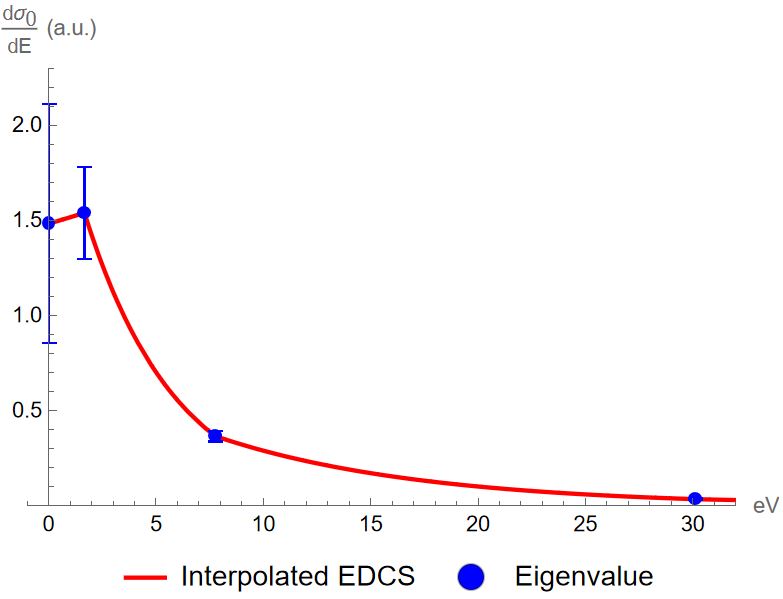}
\textbf{(a)} $l=0$
\end{minipage}
\hfill
\begin{minipage}{0.48\textwidth}
\centering
\includegraphics[width=\linewidth]{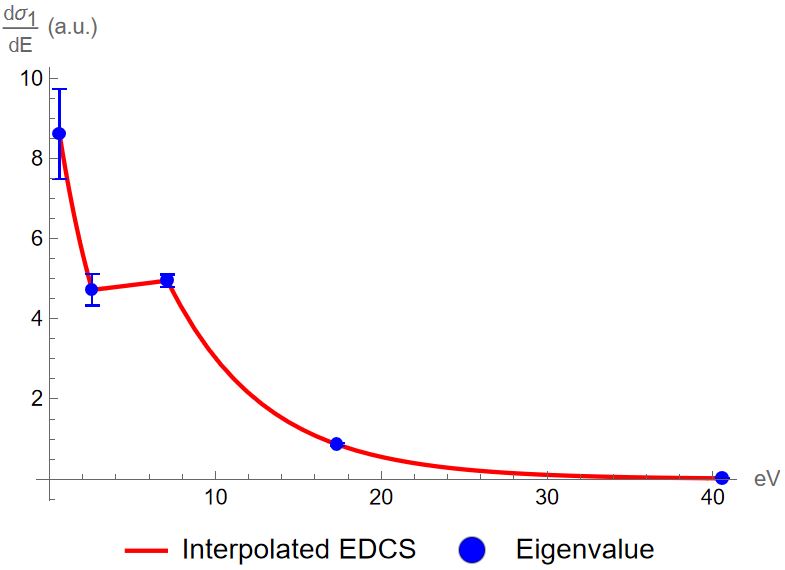}
\textbf{(b)} $l=1$
\end{minipage}

\vspace{0.5cm}

\begin{minipage}{0.48\textwidth}
\centering
\includegraphics[width=\linewidth]{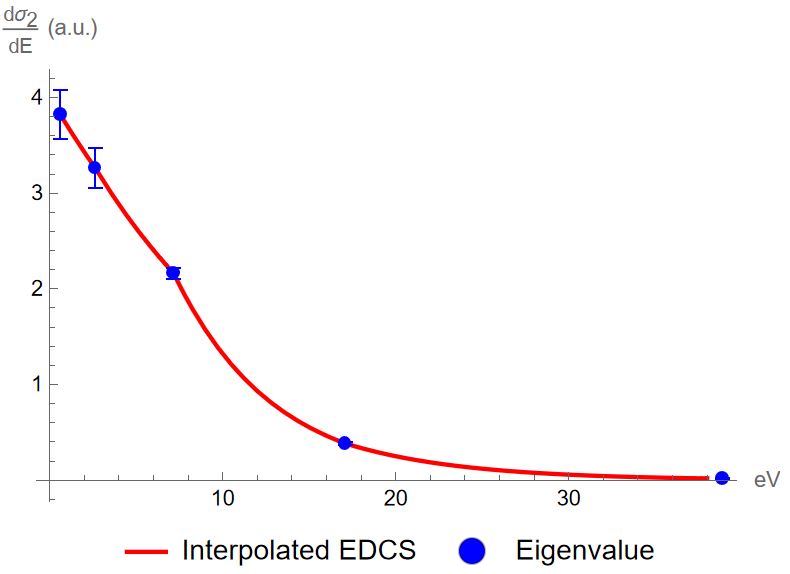}
\textbf{(c)} $l=2$
\end{minipage}
\hfill
\begin{minipage}{0.48\textwidth}
\centering
\includegraphics[width=\linewidth]{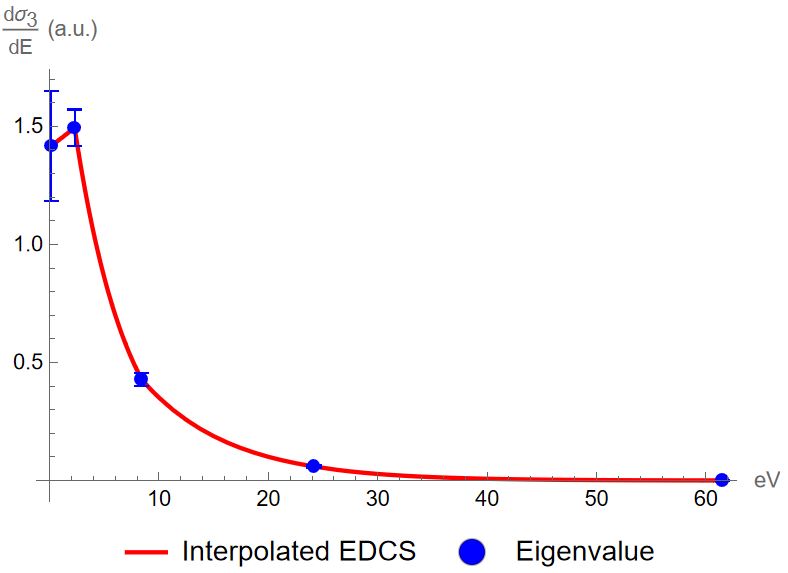}
\textbf{(d)} $l=3$
\end{minipage}

\caption{Same as Fig.\ref{fig.EDCS30} but for an antiproton impact energy of
10~keV.}
\end{figure*}

\end{document}